\documentclass[pra,twocolumn,showpacs,superscriptaddress,amssymb]{revtex4}

\usepackage[colorlinks=true,linkcolor=magenta,citecolor=magenta,urlcolor=magenta,linktocpage=true]{hyperref}
\usepackage[utf8]{inputenc}
\usepackage[english]{babel}
\usepackage[T1]{fontenc}
\usepackage{amsmath}
\usepackage{cleveref}
\usepackage{svg}
\usepackage{tabularx}
\usepackage{array, makecell}

\usepackage{listings}
\usepackage{xcolor}

\definecolor{codegreen}{rgb}{0,0.6,0}
\definecolor{codegray}{rgb}{0.5,0.5,0.5}
\definecolor{codepurple}{rgb}{0.58,0,0.82}
\definecolor{backcolour}{rgb}{0.95,0.95,0.92}
\lstdefinestyle{mystyle}{
    backgroundcolor=\color{backcolour},   
    commentstyle=\color{codegreen},
    keywordstyle=\color{magenta},
    numberstyle=\tiny\color{codegray},
    stringstyle=\color{codepurple},
    basicstyle=\ttfamily\footnotesize,
    breakatwhitespace=false,         
    breaklines=true,                 
    captionpos=b,                    
    keepspaces=true,                 
    numbers=none,                    
    numbersep=5pt,                  
    showspaces=false,                
    showstringspaces=false,
    showtabs=false,                  
    tabsize=2
}
\usepackage{psfrag,graphicx}
\usepackage{dcolumn}
\usepackage{bm}
\usepackage{amsfonts,amssymb,amsmath}        
\usepackage{slashed}
\usepackage[utf8]{inputenc}
\usepackage{graphicx}
\usepackage[caption=false]{subfig}
\usepackage{cancel}
\usepackage{xcolor}
\usepackage{amsmath}
\setcitestyle{numbers,square}

\newcommand{\figref}[1]{\mbox{Fig.~\ref{#1}}}
\newcommand{\ex}[1]{\langle #1 \rangle}

\newcommand{\be}{\begin{equation}}
\newcommand{\ee}{\end{equation}}
\newcommand{\bq}{\begin{eqnarray}}
\newcommand{\eq}{\end{eqnarray}}

\makeatletter

\renewenvironment{widetext@grid}{%
  \par\ignorespaces
  \setbox\widetext@top\vbox{%
   \vskip15\p@
   \hb@xt@\hsize{%
    \leaders\hrule\hfil
    \vrule\@height6\p@
   }%
   \vskip6\p@
  }%
  \setbox\widetext@bot\hb@xt@\hsize{%
    \vrule\@depth6\p@
    \leaders\hrule\hfil
  }%
  \onecolumngrid

  \let\set@footnotewidth\set@footnotewidth@ii
}{%
  \par

  \twocolumngrid\global\@ignoretrue
  \@endpetrue
}%

\makeatother
\begin{document}
\title{A pseudo-fermion method for the exact description of fermionic environments: \\from single-molecule electronics to Kondo resonance}

\author{Mauro Cirio}
\email{cirio.mauro@gmail.com}
\affiliation{Graduate School of China Academy of Engineering Physics, Haidian District, Beijing, 100193, China}
\author{Neill Lambert}
\email{nwlambert@gmail.com}
\affiliation{Theoretical Quantum Physics Laboratory, Cluster for Pioneering Research, RIKEN, Wakoshi, Saitama 351-0198, Japan}
\author{Pengfei Liang}
\affiliation{Graduate School of China Academy of Engineering Physics, Haidian District, Beijing, 100193, China}
\author{Po-Chen Kuo}
\affiliation{Department of Physics, National Cheng Kung University, 701 Tainan, Taiwan}
\affiliation{Center for Quantum Frontiers of Research \& Technology, NCKU, 70101 Tainan, Taiwan}
\author{Yueh-Nan Chen}
\affiliation{Department of Physics, National Cheng Kung University, 701 Tainan, Taiwan}
\affiliation{Center for Quantum Frontiers of Research \& Technology, NCKU, 70101 Tainan, Taiwan}
\author{Paul Menczel}
\affiliation{Theoretical Quantum Physics Laboratory,  Cluster for Pioneering Research, RIKEN, Wakoshi, Saitama 351-0198, Japan}
\author{Ken Funo}
\affiliation{Theoretical Quantum Physics Laboratory, Cluster for Pioneering Research, RIKEN, Wakoshi, Saitama 351-0198, Japan}
\author{Franco Nori}
\affiliation{Theoretical Quantum Physics Laboratory, Cluster for Pioneering Research, RIKEN, Wakoshi, Saitama 351-0198, Japan}
\affiliation{Center for Quantum Computing, RIKEN, Wakoshi, Saitama 351-0198, Japan}
\affiliation{Physics Department, The University of Michigan, Ann Arbor, Michigan 48109-1040, USA}

\date{\today}

\begin{abstract}
We develop a discrete fermion approach for modeling the strong interaction of an arbitrary system interacting with continuum electronic reservoirs. The approach is based on a pseudo-fermion decomposition of the continuum bath correlation functions, and is only limited by the accuracy of this decomposition. We show that to obtain this decomposition one can allow for imaginary pseudo-fermion parameters, and strong damping in individual pseudo-fermions, without introducing unwanted approximations.  For a non-interacting single-resonant level, we benchmark our approach against an analytical solution and an exact hierachical-equations-of-motion approach.  We also show that, for the interacting case,  this simple method can capture the strongly correlated low-temperature physics of Kondo resonance,  even in the difficult scaling limit, by employing matrix product state techniques.
\end{abstract}

\maketitle
\emph{Introduction}.-- The orthodox model of strongly correlated electron transport in mesoscopic physics \cite{Mahan,Zagoskin,Shevchenko} imagines one or more electronic reservoirs, parameterized by their spectral density, temperature, and chemical potential, coupled to an arbitrary system (typically fermionic impurities interacting with each other, or with additional, e.g., bosonic, environments). While this model can be often well described by \emph{weak}-coupling theories, strong coupling between system and reservoirs plays an important role in molecular electronics~\cite{molecules}, mesoscopic transport through quantum dots (artificial molecules) \cite{RevModPhys.75.1,PhysRevB.78.235311, Jin_Matisse, Matisse_arxiv}, and quantum thermodynamics \cite{PhysRevE.76.031105,thermodots,thermomoles, PhysRevX.10.031040}.  While some integrable limits exist \cite{Hewson,brandes,March}, generally speaking one must resort to numerical approaches to capture the non-trivial correlations that can build up between system and reservoirs. 

Recently several such approaches based on {\em discrete} fermion methods have been proposed and studied to capture this \emph{strong}-coupling regime \cite{Petruccione, Gardiner, PhysRevLett.109.170402}, partially inspired by similar techniques developed for strong coupling to bosonic environments \cite{Tanimura_1,PhysRevA.55.2290,iles2014environmental,Chen2015,PhysRevLett.120.030402,Lemmer_2018,Lambert,PhysRevA.101.052108}. The hierarchical equations of motion (HEOM) \cite{Tanimura_3,Ishizaki_1,Ishizaki_2,Tanimura_2014,Fruchtman,Tanimura_2020,Tanimura_2021} (which non-perturbatively accounts for system-bath entanglement by evolving a hierarchy of time-local equations obtained by repeatedly differentiating the system path-integral representation) have found great success for fermionic systems \cite{Yan_5,KondoPRL,schinabeck2,schinabeck, Lambert_Bofin,Jakub2020} and will be used as a benchmark of our results in this work (see also related stochastic methods in \cite{Tanimura_2,stoch1, stoch2}). More recently, the reaction-coordinate method \cite{PhysRevB.97.205405}, which non-perturbatively models the most relevant degrees of freedom of the environment, was adapted to fermionic systems, and is simple and transparent, both conceptually and in terms of implementation.  However, it is arguably limited as the approximations needed to treat the residual bath break down in the wide-band limit.  Similarly, a recent approach \cite{PhysRevX.10.031040} based on fitting the power spectrum with discrete Lorentzians shows promising results in terms of convergence, but, by relying on only physical modes, it inherits limitations on the width and positivity of the Lorentzian fitting functions. Finally, other methods \cite{PhysRevLett.110.086403,PhysRevB.89.165105,PhysRevB.92.245125,PhysRevB.92.125145,Dorda,PhysRevB.101.165132,ace} build up a continuum reservoir 
from a finite set of discrete damped physical modes, and can be combined with tensor network techniques (such as the process-tensor \cite{PhysRevA.97.012127,tempo,tempoadd}) for efficient construction and time-evolution of system properties.

 Here, we develop a ``pseudo-fermion method'' (akin to bosonic pseudomodes \cite{PhysRevA.55.2290,PhysRevLett.120.030402,Lemmer_2018,PhysRevA.101.052108,PhysRevResearch.2.043058}) based on a discrete set of effective fermions which reproduce the key features of the correlation functions of the continuum bath. By employing  unphysical modes with complex couplings, we model the reduced dynamics of the system by solving a Lindblad master equation in the augmented system+pseudo-fermion space. The use of a Lindblad form for the damping of the pseudo-fermions does not, by construction, introduce any approximation. The only approximation arises from how well the total correlation function of the pseudo-fermions matches that of the original bath. In contrast to other methods, \cite{Garg, Martinazzo,iles2014environmental,Woods,Strasberg_2016,Melina,Chin_2010,PhysRevLett.105.050404,PhysRevLett.123.090402,PhysRevB.101.155134}, these unphysical degrees of freedom do not have direct connection to the original physical environment allowing optimization of the modeling over an enlarged domain. In summary, we present a  conceptually simple framework to simulate non-perturbative effects in Fermionic systems by using a Lindblad master equation. To show that this simplicity does not necessarily come at the expense of modeling power, we benchmark the accuracy of the method by reproducing non-equilibrium and Kondo-physics effects.  We note that here the term pseudo-fermion has a different meaning than  in non-Hermitian quantum mechanics \cite{PhysRevLett.80.5243,Trifonov,Bagarello1,Bagarello2}.

\begin{figure}[t!]
\includegraphics[width = \columnwidth]{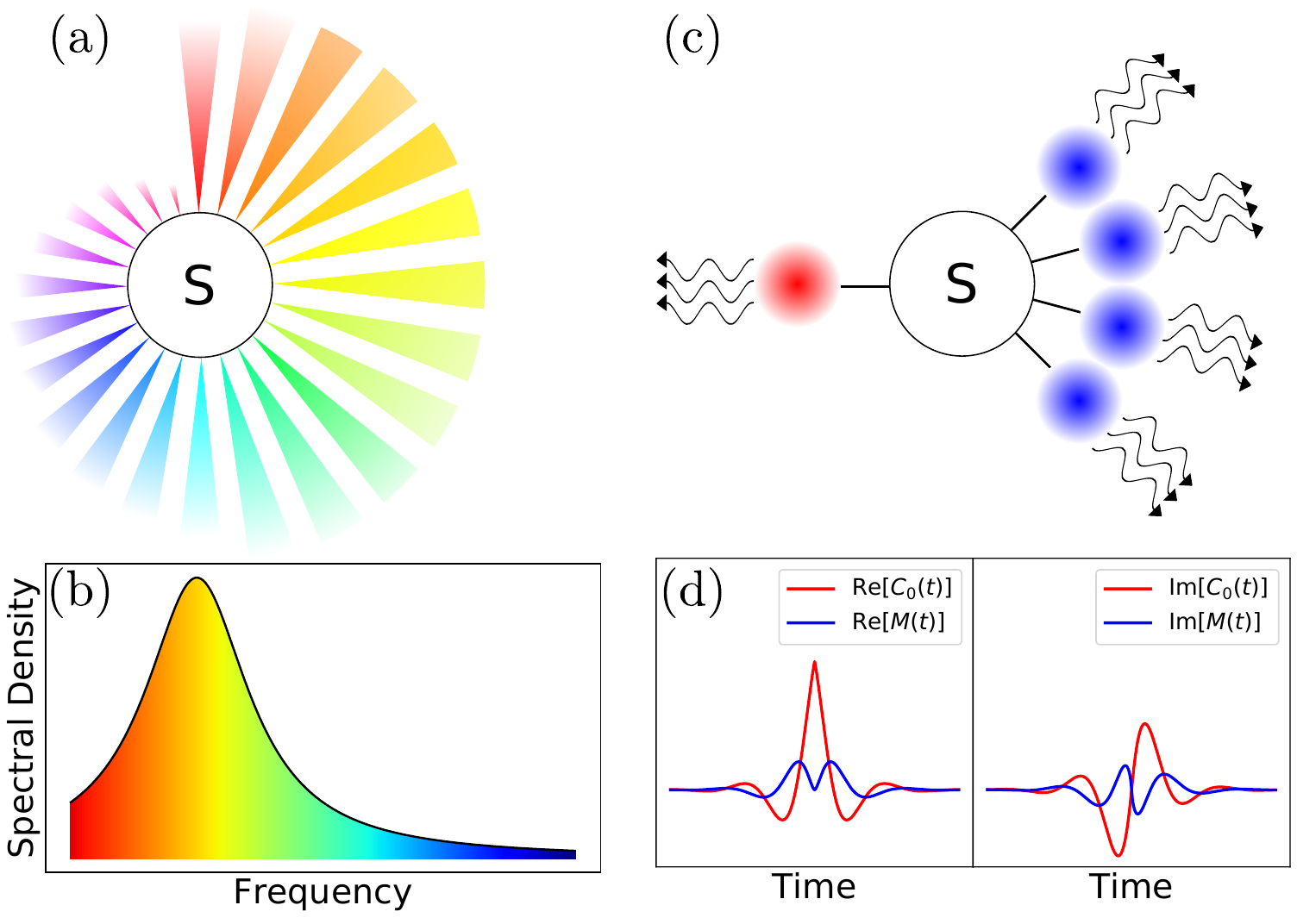}
\caption{Illustration of the pseudo-fermion model. In (a) a Fermionic system $S$ interacts with a continuum of environmental Fermionic modes at different frequencies. The spectral density function in (b) describes the distribution of the system-environment coupling. The dynamics of $S$ can be equivalently computed by considering the interaction with a discrete set of pseudo-fermions (red/blue circles) whose dissipative properties are described by a Lindblad master equation (wiggly arrows). This equivalence holds as the pseudo-fermions are designed to reproduce the correlation functions of the original bath. Specifically, for the spectral density in (b), the red (blue) pseudo-fermions in (c) reproduce the resonant (Matsubara) contribution to the correlation function plotted in (d), see Eq.~(\ref{eq:res_and_Mats}).}\label{fig:cartoon}
\end{figure}
\emph{Fermionic Open Quantum Systems}.-- We consider a Fermionic system $S$ interacting with a Fermionic Gaussian environment $E$ described by the Hamiltonian $H = H_S + H_E + H_I$ ($\hbar=1$),
where $H_S$ and  $H_E=\sum_k \omega_k c_k^\dagger c_k$ are the Hamiltonians for the system and environment (so that each environmental Fermion $c_k$ is characterized by the frequency $\omega_k$).  The interaction Hamiltonian is defined as
$H_{I}=\sum_{k} g_k(s c_{k}^{\dagger} + c_{k}s^{\dagger})$, where $g_{k}$ are the coupling strengths and $s$ is an odd-Fermion parity operator with support on the system. We assume the bath to be initially in a thermal equilibrium state $\rho_E^\text{eq}\propto\exp[-\beta \sum_k(\omega_k-\mu)c_k^\dagger c_k]$, with inverse temperature $\beta$ and chemical potential $\mu$. 

In this context, the reduced system dynamics depends, in general, on the free-bath statistical properties encoded in correlation functions involving the fields ${B}(t)=\sum_{k}g_{k}c_{k}e^{-i\omega_{k}t}$ in the interaction picture. When the free statistics is Gaussian, this dependence can be reduced down to two-point correlations, invoking Wick's theorem to write \cite{Cirio2021}
\begin{equation}
\label{eq:rhoF}
\rho_S(t)=\mathcal{T}\exp{[\mathcal{F}(t,s,C^\sigma)]}\rho_S(0)\;,
\end{equation}
in terms of the Fermionic time ordering operator $\mathcal{T}$ [see, for example, \cite{Greiner}, Eq.~(5.84)] and the Fermionic influence superoperator $\mathcal{F}(t,s,C^\sigma)$ (see Eq.~(\ref{eq:F_explicit}) in \cite{Supplemental}). This operator \emph{exactly} describes the effects of the bath on the system through its dependence on the system coupling operator $s$ and the translational-invariant two-point correlation functions $C^{\sigma}(t)=\text{Tr}_E\left[B^\sigma(t) B^{\bar{\sigma}}(0)\rho^\text{eq}_E\right]$, which explicitly read
\begin{equation}
\label{eq:corr_short}
C^{\sigma}(t)= \int_{-\infty}^\infty d\omega\;J(\omega)e^{i\sigma\omega t}[(1-\sigma)/2+\sigma n^\text{eq}_E(\omega)]/\pi\;.
\end{equation}

 Here we defined the spectral density $J(\omega)=\pi \sum_k g_k^2\delta(\omega-\omega_k)$, the Fermi distribution $n_E^\text{eq}(\omega)=(1+\exp[\beta(\omega-\mu)])^{-1}$, and $\sigma=\pm 1$ to denote the presence/absence of Hermitian conjugation, i.e., $B^{\sigma=1}=B^{\dagger}$ and $B^{\sigma=-1}=B$ (with $\bar{\sigma}=-\sigma$).

This analysis implies that \emph{all memory effects} present in non-perturbative regimes  can be encoded in the superoperator $\mathcal{F}(t,s,C^\sigma)$. Therefore, Gaussian open systems with identical environmental correlations $C^\sigma(t)$ (and same system-coupling operators $s$)  must have identical influence superoperators and, consequently, an equivalent reduced system dynamics through Eq.~(\ref{eq:rhoF}).

\emph{Pseudo-Fermion method}.-- 
We now proceed as schematically shown in Fig.~\ref{fig:cartoon}(c), i.e., instead of solving the original system + continuum environment Hamiltonian, we define a new model consisting of the original system (S) in contact with a set of discrete pseudo-fermions (pf), themselves in contact with residual environments (re). The key point is that our new environment (pf + re) is designed to produce the same correlation functions of the original environment, and hence produce the same system dynamics, as per Eq.~(\ref{eq:rhoF}). To satisfactorily mimic  the original correlation functions with only a finite discrete set of artificial systems, we allow certain parameters defining these pseudo-fermions to be unphysical. Importantly, each residual environment is an idealized quantum white noise (defined by constant spectral densities and frequency-independent equilibrium distributions), thereby allowing the dynamics in the pseudo-fermion space to be \emph{exactly} described by a simple Lindblad equation [wiggly arrows in Fig.~\ref{fig:cartoon}(c)]. Surprisingly, the use of Lindblads does not limit the accuracy of the result as the derivation does not rely on approximations as long as the correlation functions produced by said Lindblads closely match the original bath ones.   

Explicitly, we consider $N_{\text{pf}}$ pseudo-fermions $\tilde{c}_j$, $j=1,\dots,N_{\text{pf}}$ interacting with the system as described by the Hamiltonian
\begin{equation}
    H_{S+\text{pf}}=H_S+H_\text{pf}+\sum_{j=1}^{N_\text{pf}} \lambda_j(s\tilde{c}^\dagger_j+ \tilde{c}_j s^\dagger)\;,\\
\end{equation}
 where $H_\text{pf}=\sum_{j=1}^{N_{\text{pf}}}\Omega_j \tilde{c}_j^\dagger \tilde{c}_j$ is the free pseudo-fermion Hamiltonian which depends on the (formal) energies $\Omega_j\in\mathbb{C}$ and the fields $\tilde{B}^\sigma_j=\lambda_j\tilde{c}^\sigma_j$, in terms of the interaction strengths $\lambda_j\in\mathbb{C}$. We further assume the pseudo-fermions and residual environments to be initially in their equilibrium state $\rho_{\text{pf}+\text{re}}^\text{eq}$. As mentioned, the residual environment associated to each pseudofermion $j$ is modeled as quantum white noise characterized by a formal decay rate $\Gamma_j\in\mathbb{C}$ and a formal Fermi distribution $n_j\in\mathbb{C}$. As shown in \cite{Supplemental}, the free correlations in the pseudo-fermion and residual-environment space, i.e.,
\begin{equation}
\label{eq:corr_pf_free}
C^\sigma_\text{pf}(t)\equiv \sum_{j=1}^{N_\text{pf}} C^\sigma_{\text{pf},j}(t)\equiv\sum_{j=1}^{N_\text{pf}}\lambda_j^2\text{Tr}_{\text{pf}+\text{re}}[\tilde{c}^\sigma_j(t)\tilde{c}^{\bar{\sigma}}_j(0)\rho_{\text{pf}+\text{re}}^\text{eq}]\;,
\end{equation}
are defined using operators in the interaction picture and they can be obtained by directly solving the Heisenberg equation of motion leading to
\begin{equation}
\label{eq:corr_pf_pf}
\begin{array}{lll}
C^\sigma_{\text{pf},j}(t)=\lambda_j^2[(1-\sigma)/2+\sigma n_j]\exp{[i\sigma\Omega_jt-\Gamma_j |t|]}\;.
\end{array}
\end{equation}
 This result shows a main feature of the pseudo-fermion method: unphysical properties (for example, the fields $\tilde{B}^\dagger_j$ \emph{not} depending on the conjugate of the parameters $\lambda_j$) allow to model a more general set of correlation functions (for example, $C^\sigma_{\text{pf},j}(0) < 0$ requires $\lambda_j$ to be imaginary).
\begin{figure}[t!]
\includegraphics[width = \columnwidth]{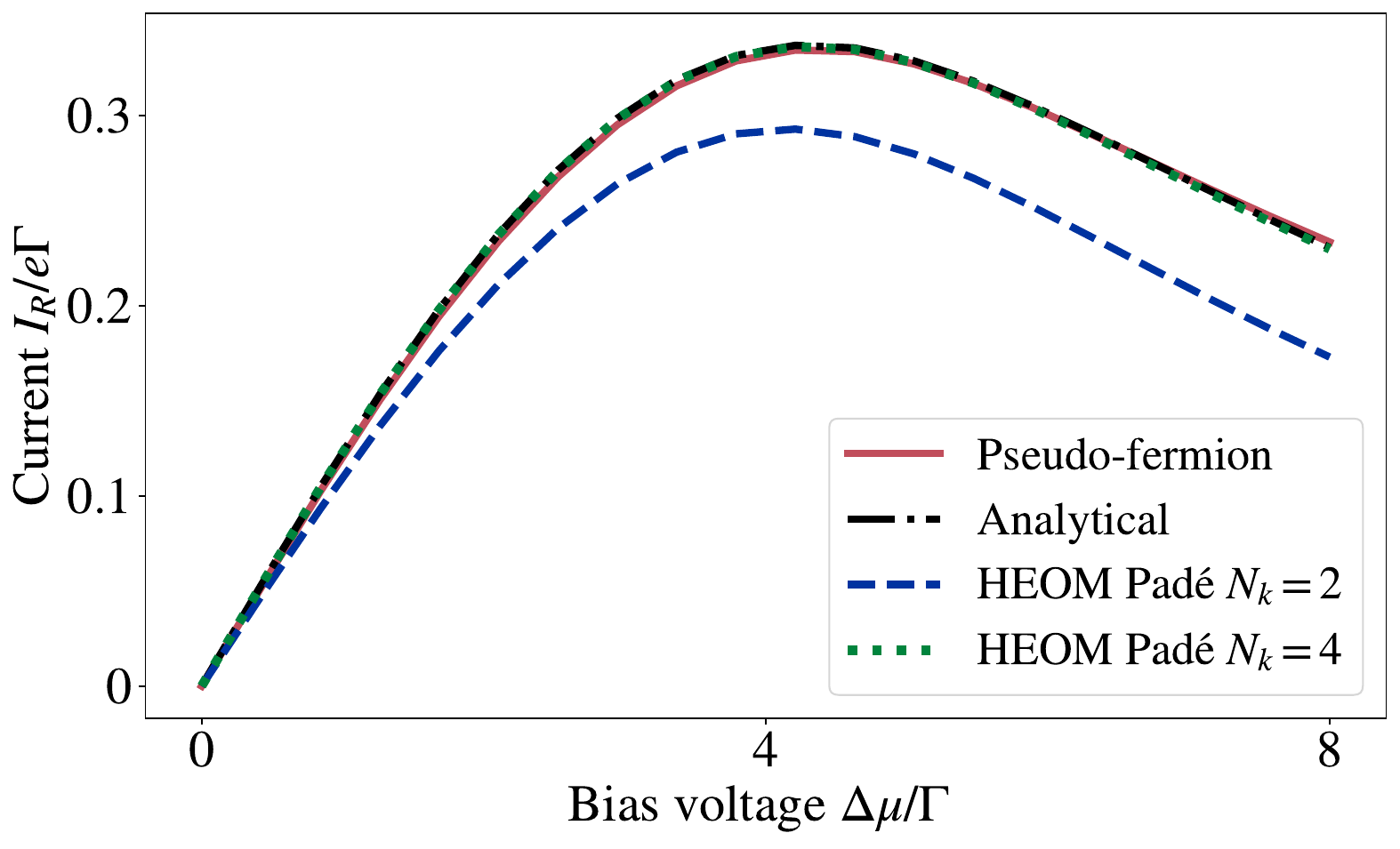}
\caption{Current into the right reservoir from a single impurity (coupled to two reservoirs) as a function of chemical potential difference $\Delta \mu= \mu_L -\mu_R$. For both reservoirs we choose $ \epsilon = \Gamma$, $W = 2.5\Gamma$, and $\beta = 1/(0.2\Gamma)$. The truncation parameter for the HEOM results is $n_{\mathrm{max}}=2$. }\label{figure2}
\end{figure}

Thanks to the Gaussianity hypothesis, the reduced dynamics of a system embedded in the environment made of pseudo-fermions and their residual environments  is equivalent to the reduced dynamics of the original model with Hamiltonian $H$ as long as
\begin{equation}
\label{eq:corr_full_pf}
C^\sigma(t)= C^\sigma_\text{pf}(t)\;.
\end{equation}
The practical advantage provided by this result is that, by choosing the residual environments to be idealized white noise, their effects on the system+pseudofermions space can be modeled (without approximations, see \cite{Cirio2021, Supplemental}) by the following Lindblad master equation 
\begin{equation}
\label{eq:master_eq_pf}
\begin{array}{lll}
\dot{\rho}_{S+\text{pf}}
&=&-i[H_{S+\text{pf}},\rho_{S+\text{pf}}]\\
&+&\displaystyle\sum_{r,j}\Gamma_j \left\{(1-n_j)D^r_{\tilde{c}_j}[\rho^r_{S+\text{pf}}]+ n_j D^r_{\tilde{c}^\dagger_j}[\rho^r_{S+\text{pf}}]\right\},
\end{array}
\end{equation}
where $r=\pm 1$ and $j=1\cdots N_{\text{pf}}$.  Here, $\rho^{r=\pm}_{S+\text{pf}}$, are the projections of the density matrix into the space with even/odd Fermionic parity and the dissipators $D_O^r[\cdot]$ are defined as $D_O^r[\cdot]=2rO[\cdot]O^\dagger-O^\dagger O[\cdot]-[\cdot]O^\dagger O$ (see also \cite{Schwarz}).
\emph{This is the main result of this article}: when the Fermionic environment of a Gaussian open quantum system has free correlations which are equivalent to (or can be approximated by) Eq.~(\ref{eq:corr_full_pf}), the reduced system dynamics can be equivalently (approximately) computed by solving the master equation in Eq.~(\ref{eq:master_eq_pf}). \\

\emph{Case study: Lorentzian spectral density}.-- We now provide an explicit pseudo-fermion model to approximate a Fermionic environment initially in an equilibrium state $\rho^\text{eq}_{E}$ with inverse temperature $\beta$ and chemical potential $\mu$, and where the interaction with the system is characterized by a Lorentzian spectral density $J_L(\omega)={\Gamma W^2}/{[(\omega-\mu)^2+W^2]}$,
where the frequency parameters $W$ and $\Gamma$  specify the width and the overall interaction strength, respectively. By inserting this expression into Eq.~(\ref{eq:corr_short}), it is possible \cite{Supplemental} to write the  decomposition: 
\begin{equation}
\label{eq:alternative_mats}
C_L^\sigma(t)=C^\sigma_\text{res}+ \sum_{k>0} M_k^\sigma(t)\;.
\end{equation}
Here, the ``resonant'' and ``Matsubara'' contributions are defined through 
\begin{equation}
\label{eq:res_and_Mats}
\begin{array}{lll}
C^{\sigma}_\text{res}(t)&=&\displaystyle \frac{\Gamma W}{2} \exp{[i \sigma\mu t - W |t|]}\\
M_k^{\sigma}(t)&=&\displaystyle M_k e^{-(W+x_k)|t|/2}\sum_{r=\pm}re^{[i\sigma\mu+r(W-x_k)/2]t},
\end{array}
\end{equation}
where $M_k={2i\Gamma W^2}/{\beta(x^2_k-W^2)}$, with $x_k=(2k-1)\pi/\beta$.
Our goal is to implement these contributions using a set of pseudo-environments  characterized by the free correlations in Eq.~(\ref{eq:corr_pf_pf}). To describe this correspondence, we  define one \emph{resonant} and two \emph{Matsubara} pseudo-environments (for each Matsubara frequency) by identifying $j\mapsto\text{res}$ and $j\mapsto (k,r=\pm)$ in Eq.~(\ref{eq:corr_pf_pf}). This leads to the following equivalences among correlations
\begin{equation}
\label{eq:equivalences}
    C^\sigma_{\text{pf},\text{res}}(t)=C^{\sigma}_\text{res}(t),~~~~~~
    \sum_{r=\pm}C^\sigma_{\text{pf},(k,r)}(t)=M_k^{\sigma}(t)\;,
    \end{equation}
which hold \cite{Supplemental} when the parameters for the pseudo-environments are defined as
\begin{equation}
\begin{array}{llllll}
n_\text{res}&=&1/2,&n_{k,\pm}&=&\Delta\\
\lambda_\text{res}&=&\sqrt{\Gamma W}, &\lambda_{k,\pm}&=&\sqrt{ \pm M_k / \Delta } \\
\Omega_\text{res}&=&\mu,&\Omega_{k,\pm}&=&\mu\mp i(x_k-W)/2\\
\Gamma_\text{res}&=&W,&\Gamma_{k,\pm}&=&(W+x_k)/2\;,
\end{array}
\end{equation}
where $\Delta\in\mathbb{C}$ with $|\Delta|\rightarrow \infty$.  We note that the limit $|\Delta|\rightarrow\infty$ might introduce numerical instabilities which can be regularized using intermediate values, such that $\Delta\gg 1$. The parameter $\Delta$ allows to match the  $\sigma$-dependence between the correlations in Eq.~(\ref{eq:res_and_Mats}) and Eq.~(\ref{eq:corr_pf_pf}), see \cite{Supplemental}. However, it is also possible to build an alternative model which does not require any asymptotic limit, but uses four pseudo-fermions for each Matsubara contribution $M_k^{\sigma}(t)$ \cite{Supplemental}.

\begin{figure*}[t!]
\includegraphics[width =\columnwidth]{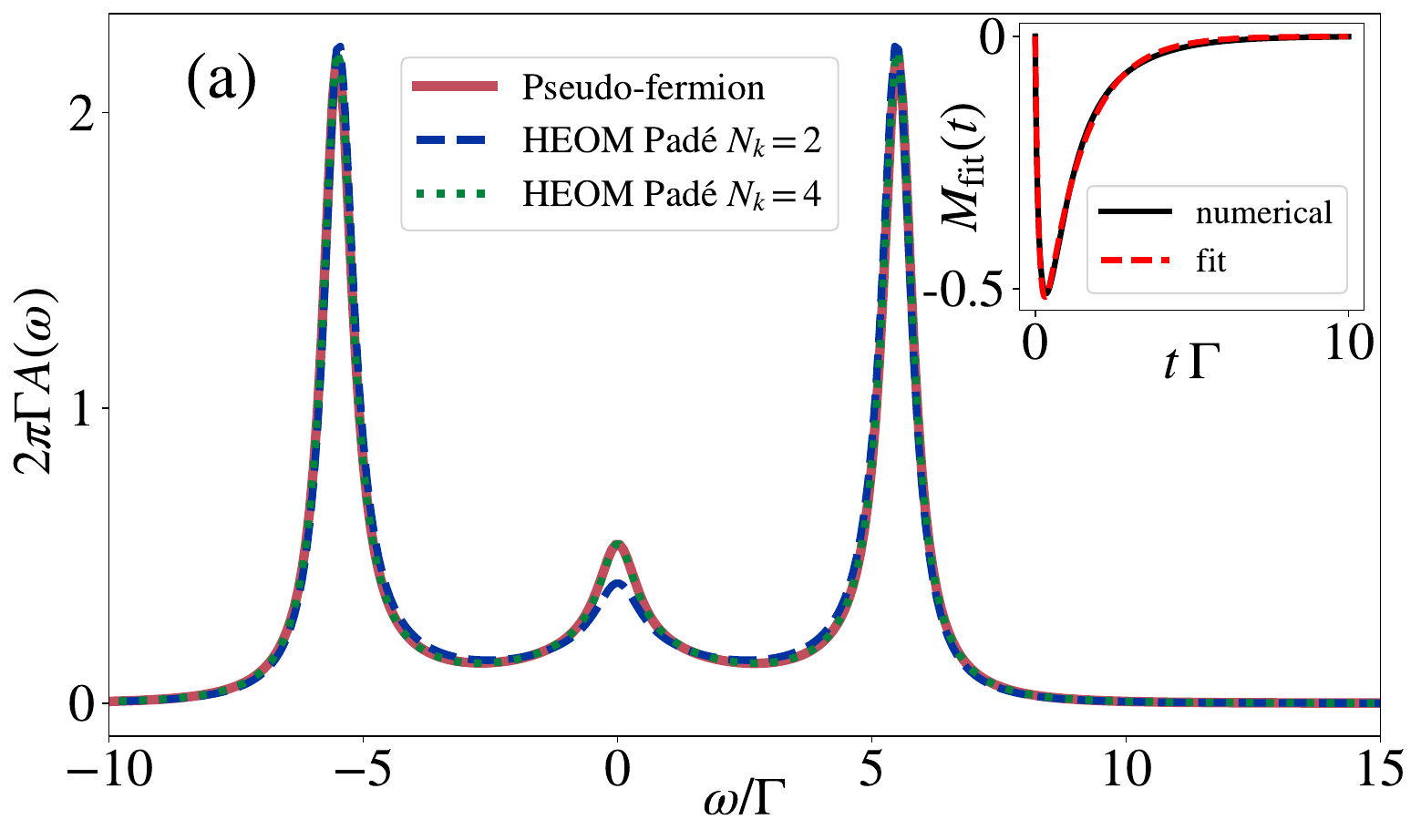}
 \includegraphics[width =\columnwidth]{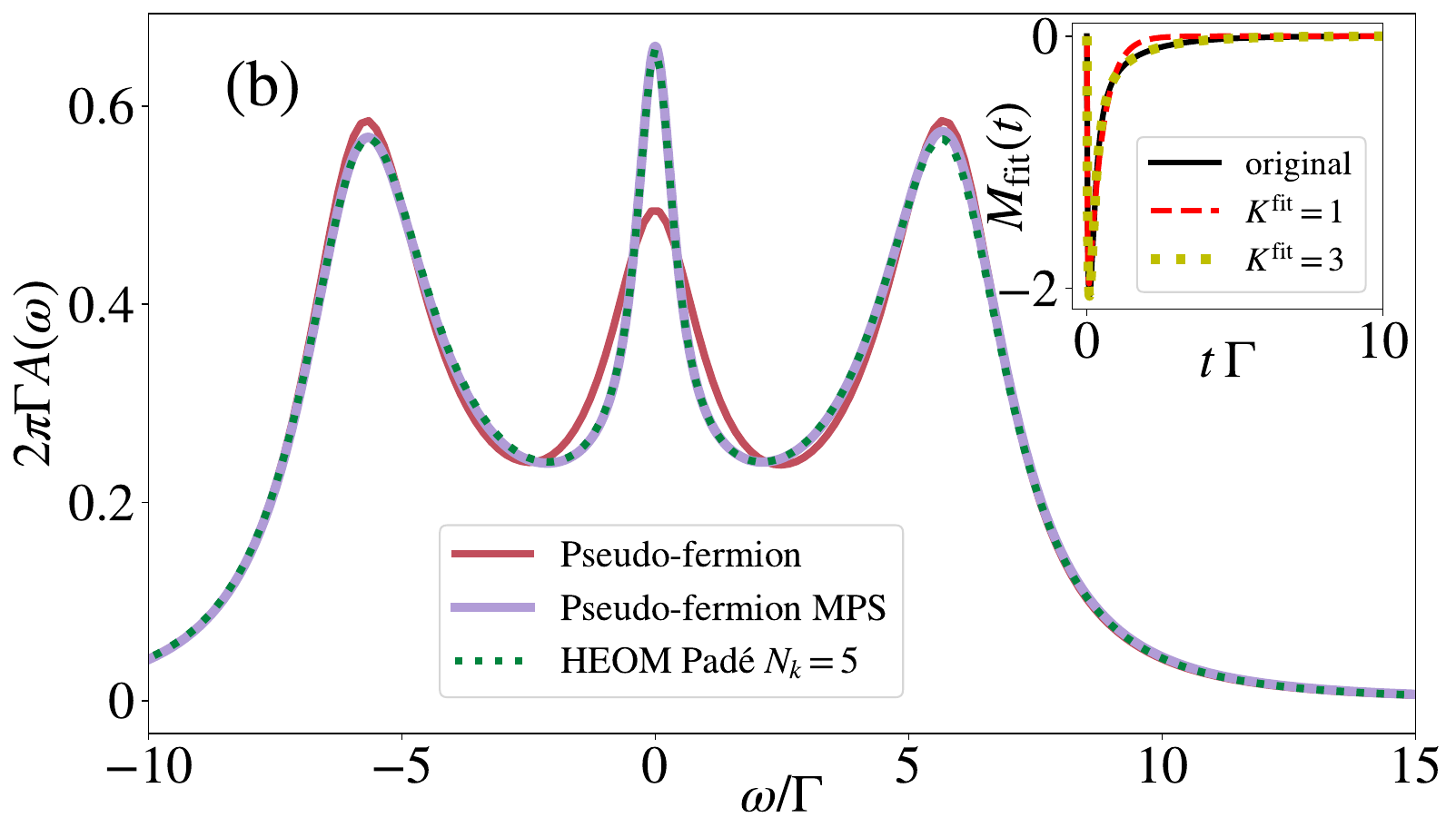}
\caption{Spectral density $A(\omega)$ for a single impurity containing two interacting fermions coupled to a spin-labelled reservoir. Here $\mu=0$, $U=3\pi \Gamma$, $\epsilon = -U/2$, and $\beta = 1/(0.2 \Gamma)$. (a) shows the results for width $W = 2.5\Gamma$, while (b) is for $W = 10\Gamma$. 
In (a) we see that the HEOM result with $N_k=2$ has not converged, while all other results are (pseudo-fermion and HEOM, $N_k=4$ results overlap). In (b), which uses the Matsubara series in Eq.~(\ref{eq:tempppp}), the HEOM results converge for $N_k=5$, while the standard pseudo-fermion results, relying on  $K^\text{fit}=1$ terms in the series in Eq.~(\ref{eq:ansatz}) to fit the Matsubara contribution (shown in the inset), have not converged.  To efficiently include more exponents, and hence more pseudo-fermions, we employ matrix product states, allowing us to include $K^\text{fit}=3$ term in the Matsubara series (yellow curve in inset) and achieve convergence (solid purple curve).
}
\label{figure3}
\end{figure*}
The mapping above is possible because of \emph{the enlarged parameter domain of the model which allows for complex (i.e., unphysical) values}. We can appreciate this by noting that (i) $M_k$ has a non-trivial overall complex phase requiring complex couplings $\lambda_{k,\pm}$ between the pseudomodes and the system, (ii) $M_k^\sigma$ contains terms proportional to $e^{\alpha t}$ ($\alpha\in\mathbb{R}$) which neither represent a pure physical oscillation ($\alpha$ is not imaginary) nor a pure physical decay ($t$ does not appear in absolute value) ultimately requiring the pseudomodes frequencies $\Omega_{k,\pm}$ to acquire an imaginary part, and (iii) the unbounded nature of the average Fermion number $n_{k,\pm}$ which implies unphysical density matrices.

The relations in Eq.~(\ref{eq:equivalences}) lead to an equivalence between the full pseudo-environment and the original model, i.e., $C^\sigma_{L}(t)=C^\sigma_{\text{pf}}(t)$. This shows that a Fermionic environment characterized by the Lorentzian spectral density $J_L(\omega)$ can be approximated using $N_\text{pf}=1+2 K_\text{pf}$ pseudo-fermions. Here $K_\text{pf}$ represents a cut-off in the evaluation of the Matsubara series in Eq.~(\ref{eq:alternative_mats}), i.e., $\sum_{k>0}\rightarrow\sum_{k=1}^{K_\text{pf}}$. 
However, when several Matsubara frequencies effectively contribute to the original correlation $C_L^\sigma(t)$ (i.e., $K_\text{pf}\gg 1$), it might be more efficient to alternatively finding a best-fit approximation of the full Matsubara series $M^\sigma(t)\equiv\sum_{k>0} M_k^\sigma(t)$, following the ansatz $M^\sigma\simeq \sum_{j=1}^{K^\text{fit}}M^\sigma_{j,\text{fit}}(t)$, where  $K^\text{fit}\in\mathbb{N}$ and
\begin{equation}
\label{eq:ansatz}
M^\sigma_{j,\text{fit}}(t)\simeq M_{j,\text{fit}}e^{-(W^j_\text{fit}+x^j_\text{fit})|t|/2}\sum_{r=\pm}re^{[i\sigma\mu+r(W^j_\text{fit}-x^j_\text{fit})/2]t},
\end{equation}
and only then proceeding with the pseudo-fermion mapping, thereby optimizing the number $N_\text{pf}^\text{fit}=1+2 K^\text{fit}$ of pseudo-fermions.

\emph{ Non-equilibrium Single Impurity Model.--} Our first example is a single spin-less fermion coupled to two environments with Hamiltonian $H=H_S+H_E+H_I$, where
\begin{equation}
\begin{array}{lll}
H_S &=& \epsilon s^{\dagger}s\\
H_E+H_I &=& \sum_{\alpha,k}\omega_{\alpha,k} c^{\dagger}_{\alpha,k}c_{\alpha,k}+\sum_\alpha (sB_\alpha^{\dagger} + B_\alpha s^{\dagger})\;,
\end{array}
\end{equation}
with $B_\alpha= \sum_{k}g_{\alpha,k} c_{\alpha,k}$ for $\alpha\in \{L,R\}$, indexing the `left' and `right' leads. We choose both environments to be identical apart from their chemical potential, so that $J_L(\omega)$ is generalized to have an $\alpha$-dependent $\mu_\alpha$, and we define $\Delta \mu = \mu_L-\mu_R$.  We then apply the fitting procedure described in Eq.~(\ref{eq:ansatz}), so that each environment is described by $N^{\text{fit}}_{\mathrm{pf}}=1+2$ pseudo-fermions. As a benchmark we use the standard analytical result (see \cite{brandes}) for the current, and the HEOM method using the BoFiN-HEOM package \cite{Lambert_Bofin} with a Pade decomposition of the bath exponents.  

For the pseudo-fermions, we can evaluate the current following the logic in \cite{schinabeck,schinabeck2,phdthesis}, defining the occupation of bath $\alpha$ as $N_\alpha=\sum_{k} c^{\dagger}_{\alpha,k}c_{\alpha,k}$, and the current into the bath $\alpha$ as $I_\alpha(t)=e\mathrm{Tr}[N_\alpha \dot{\rho}(t)] =-i e\mathrm{Tr}[\{s(t) B_\alpha^{\dagger}(t) + s^{\dagger}(t)B_\alpha(t)\}\rho(t)]$.  In the pseudo-fermion formalism we equate $B_\alpha$ with a sum over all pseudo-fermions describing environment $\alpha$.  Equivalence of the results \cite{Qutip1,Qutip2} for the analytical current, the HEOM method, and the pseudo-fermions method is demonstrated in \figref{figure2}.

\emph{Kondo resonance.--} To demonstrate that the pseudo-fermions can indeed capture non-trivial correlations between system and environment we turn to the example of Kondo resonance~\cite{PhysRevLett.121.026805,PhysRevB.98.024103,Ivan2020,kondocloud,kondocorrelation}. Previously, this was used to demonstrate the power of the HEOM method in dealing with Fermionic systems~\cite{KondoPRL}, and here we do the same for the pseudo-fermion method.  We start with a system containing two interacting Fermions labelled by their spin coupled to a spin-labelled reservoir,
\begin{equation}
    \begin{array}{lll}
H_S &=&  \epsilon\left(s_{\uparrow}^{\dagger}s_{\uparrow} + s_{\downarrow}^{\dagger}s_{\downarrow}\right) + Us_{\uparrow}^{\dagger}s_{\uparrow} s_{\downarrow}^{\dagger}s_{\downarrow}\\
H_E + H_I &=& \sum_{k,\nu} c^{\dagger}_{k,\nu}c_{k,\nu}+\sum_{\nu} s_{\nu}B_{\nu}^{\dagger} + B_{\nu}s_{\nu}^{\dagger}\;,
\end{array}
\end{equation}
where $\nu \in \{\uparrow, \downarrow\}$. The spectral density of the impurity with spin $\nu$ (an experimentally observable quantity~\cite{Angleresolved,STM}  which can exhibit signatures of Kondo resonance) is defined as $A_{\nu}(\omega) = \int dt e^{i\omega t} \ex{\{s_{\nu}(t),s_{\nu}^{\dagger}(0)\}}/2\pi$.
This can be evaluated using the pseudo-fermion equation of motion. \figref{figure3} demonstrates the spectrum for a symmetric example where $U = -2 \epsilon $.  The pseudo-fermion method fits the predicted HEOM result  (using a converged Pad\'e decomposition of the bath correlation functions) remarkably well.  The two side-peaks appear around the system energies, while the Kondo resonance appears at zero frequency, as expected. 

Importantly, by considering the full Matsubara series $M^\sigma(t)$, we show in  \cite{Supplemental} that all terms associated with large (compared to the inverse time-scales of the system) Matsubara frequencies \emph{do not} effectively contribute to the reduced system dynamics (while for Bosonic environments such terms introduce additional Markovian decay \cite{Ishizaki_1}). This implies that a finite truncation of the series is possible, and hence a finite number of pseudomodes can be employed even in the scaling limit. However, despite this, direct numerical solutions using standard Jordan-Wigner mappings become numerically challenging when more than a few pseudomodes are needed.  Fortunately, by employing matrix product state techniques, akin to those used in \cite{PhysRevX.10.031040,PhysRevLett.123.100502}, we have found that we can access this regime, as demonstrated in Fig.~\ref{figure3}(b)(see also \cite{Supplemental}).

\emph{Summary.--} We presented a methodology to generate the reduced dynamics of an arbitrary system interacting with Fermionic environments beyond Born-Markov approximations. The model parameters can break certain physical constraints thereby allowing for greater possibility of optimization. This method allows to accurately simulate challenging non-Markovian regimes by solving a Lindblad master equation without resorting to approximations as long as the correlation functions closely match the ones of the original bath. We demonstrated this balance between conceptual simplicity and numerical accuracy by reproducing the exact results of both non-equilibrium transport and Kondo resonance by benchmarking against the HEOM method.  The latter is known  \cite{KondoPRL} to perform as well as renormalization-group methods at finite temperatures, and out-perform continous-time 
quantum Monte Carlo, Green's function equations of motion, and slave-boson mean-field theory. It is important to iterate that, compared to the powerful HEOM method, the pseudofermion method produces identical results, with the advantage of a more transparent interpretation (discrete effective bath fermions) which, as we have demonstrated, are then more intuitively amenable to MPS techniques \cite{HEOMMPS2,HEOMMPS1}. 

In summary, the method can be applied to simulate the effects of any fermionic environment on a quantum system with a Lindblad master equation thereby allowing challenging regimes to become more accessible by a broader audience. This could allow to design control/environmental-engineering protocols, 
to study superconducting/Majorana leads or hybrid environments (involving both fermions and bosons).  Another interesting future direction is a generalization to compute bath observables and correlations to study non-equilibrium physics within the environment itself, possibly leading to optimize transport properties.

\section{Acknowledgements}
We acknowledge Russell Deacon and Xiao Zheng for discussions and feedback.  M.C. acknowledges support from NSFC (Grants No.~12050410264 and No.~11935012) and NSAF (Grant No.~U1930403). N.L.~acknowledges partial support from JST PRESTO through Grant No.~JPMJPR18GC, and the Information Systems Division, RIKEN, for use of their facilities. F.N. is supported in part by: Nippon Telegraph and Telephone Corporation (NTT) Research, the Japan Science and Technology Agency (JST) [via the Quantum Leap Flagship Program (Q-LEAP), and the Moonshot R\&D Grant Number JPMJMS2061], the Japan Society for the Promotion of Science (JSPS) [via the Grants-in-Aid for Scientific Research (KAKENHI) Grant No. JP20H00134], the Army Research Office (ARO) (Grant No. W911NF-18-1-0358), the Asian Office of Aerospace Research and Development (AOARD) (via Grant No. FA2386-20-1-4069), and the Foundational Questions Institute Fund (FQXi) via Grant No. FQXi-IAF19-06. F.N., N.L.~and P.M.~acknowledge the Foundational Questions Institute Fund (FQXi) via Grant No.~FQXi-IAF19-06. YNC acknowledges the support of the Ministry of Science and Technology, Taiwan (MOST Grants No. 111-2123-M-006-001).
\begin{widetext}
\begin{center}

\newpage
\textbf{\large Supplemental Material to: \\A pseudo-fermion method for the exact description of fermionic environments:\\ from single-molecule electronics to Kondo resonance}\\
\vspace{.3cm}
Mauro Cirio, Neill Lambert, Po-Chen Kuo,  Yueh-Nan Chen, Paul Menczel, Ken Funo, and Franco Nori\\
\vspace{.5cm}
In this supplemental material, we present technical details supporting the main text.
\vspace{.5cm}
\end{center}

 \end{widetext}
 
 \setcounter{equation}{0}
\setcounter{figure}{0}
\setcounter{section}{0}
\setcounter{table}{0}
\setcounter{page}{1}
\makeatletter

\appendix
\section{Fermionic Open Quantum System}
In this section, we present technical details about the Fermionic open quantum systems described by the original System-Environment Hamiltonian $H=H_S+H_E+H_I$ used in the main text. Under the hypothesis described in the main article, the reduced system dynamics depends solely on the influence superoperator presented in Eq.~(\ref{eq:rhoF}) which can be written as
\begin{equation}
\label{eq:F_explicit}
\mathcal{F}(t)=\sum_{\sigma=\pm}\int_0^t dt_2\int_0^{t_2}dt_1\mathcal{A}^\sigma(t_2)\mathcal{B}^\sigma(t_2,t_1)\;,
\end{equation}
where
\begin{equation}
\label{eq:AB}
\begin{array}{lll}
\mathcal{A}^\sigma(t)[\cdot]&=&s^{\bar{\sigma}}(t)[\cdot]-\mathcal{P}[\cdot s^{\bar{\sigma}}(t)]\\
\mathcal{B}^\sigma(t_2,t_1)[\cdot]&=&-C^\sigma_{21}s^\sigma(t_1)[\cdot]-\bar{C}^{\bar{\sigma}}_{21}\mathcal{P}[\cdot s^\sigma(t_1)]\;.
\end{array}
\end{equation}
Here we defined $s^{\sigma=-1}\equiv s$ and  $s^{\sigma=1}\equiv s^\dagger$. The parity operator is defined as
\begin{equation*}
    \mathcal{P}[\cdot]=P_S[\cdot]P_S\;,
\end{equation*} 
where 
\begin{equation*}
    P_S=\prod_{k_S}\exp\left\{i\pi d_{k_S}^\dagger d_{k_S}\right\}\;,
\end{equation*}
in terms of the Fermions $d_{k_S}$ which can populate the system. The expression of the influence superoperator explicitly shows us that the effects of the environment are fully encoded in the correlation functions $C^\sigma_{21}\equiv C^\sigma(t_2,t_1)$, where
\begin{equation}
\label{eq:corr}
\begin{array}{lll}
C^{\sigma=1}(t_2,t_1)&=&\displaystyle\text{Tr}_E\left[B^\dagger(t_2)B(t_1)\rho_E^\text{eq}\right]\\
&=&\displaystyle\frac{1}{\pi}\int_{-\infty}^\infty d\omega J(\omega) n_E^\text{eq}(\omega) e^{i\omega (t_2-t_1)}\\
C^{\sigma=-1}(t_2,t_1)&=&\displaystyle\text{Tr}_E\left[B(t_2)B^\dagger(t_1)\rho_E^\text{eq}\right]\\
&=&\displaystyle\frac{1}{\pi}\int_{-\infty}^\infty d\omega J(\omega) [1-n_E^\text{eq}(\omega)] e^{-i\omega (t_2-t_1)},
\end{array}
\end{equation}
in terms of the interaction picture operators ${B}(t)=\sum_k g_k c_k e^{-i\omega_k t}$, the spectral density $J(\omega)=\pi \sum_k g_k^2\delta(\omega-\omega_k)$, and the Fermi distribution $n_E^\text{eq}(\omega)=(1+\exp\{\beta(\omega-\mu)\})^{-1}$ (in which $\mu$ represents the chemical potential). These definitions are the explicit version of  the ones in  Eq.~(\ref{eq:corr_short}) upon noticing that the invariance of $\rho^\text{eq}_E$ under the free evolution of the bath implies time-translational invariance for the correlations.
\section{Pseudofermion Model}
\label{app:PF_model}
In this section we present details about the pseudo-fermion model presented in the main text.
The model is an approximation scheme which operates in  an augmented Hilbert space involving ancillary pseudo-fermions. The prefix pseudo highlights the absence of physicality constraints other than the requirement to generate the correct reduced system dynamics. Explicitly, we consider $N_{\text{pf}}$ pseudo-fermions $\tilde{c}_j$, $j=1,\dots,N_{\text{pf}}$, each interacting with its own residual bath of Fermions $\tilde{c}_{jk}$ defining a formal open quantum system with Hamiltonian $H^\text{tot}_{S+\text{pf}}=H_{S+\text{pf}}+H_{RE}$. Here, $H_{S+\text{pf}}=H_S+H_\text{pf}+H^I_{S+\text{pf}}$ is the free Hamiltonian in the augmented system+pseudo-fermions space, where $H_\text{pf}=\sum_{j=1}^{N_{\text{pf}}}\Omega_j \tilde{c}_j^\dagger \tilde{c}_j$ is the free pseudo-fermion Hamiltonian which depends on the (formal) energies $\Omega_j\in\mathbb{C}$ and where
\begin{equation}
\begin{array}{lll}
H^I_{S+\text{pf}}&=&\sum_{j=1}^{N_\text{pf}} \lambda_j(s\tilde{c}^\dagger_j+ \tilde{c}_j s^\dagger)\equiv \sum_{j=1}^{N_\text{pf}} (s \tilde{B}_j^\dagger + \tilde{B}_j s^\dagger)\\
H_{RE}&=&\displaystyle\sum_{j,k}^{N_{\text{pf}}} \Bigl[
\displaystyle\Omega_{jk} \tilde{c}^\dagger_{jk} \tilde{c}_{jk}+\lambda_{jk}(\tilde{c}_j^\dagger \tilde{c}_{jk}+ \tilde{c}^\dagger_{jk}\tilde{c}_j) \Bigr]
\end{array}
\end{equation}
are the system+pseudo-fermions and pseudo-fermions+residual-environment interaction Hamiltonians written in terms of the fields $\tilde{B}^\sigma_j(t)=\lambda_j\tilde{c}^\sigma_j(t)$, the energies $\Omega_{jk}\in\mathbb{R}$, and the interaction strengths $\lambda_j\in\mathbb{C}$, $\lambda_{jk}\in\mathbb{R}$.
We further assume each auxiliary environment to act as an idealized white noise defined by constant spectral densities $J_j(\omega)=\pi\sum_{k}\lambda^2_{jk}\delta(\omega-\Omega_{jk})=\Gamma_j$ (in terms of rates $\Gamma_j\in\mathbb{C}$) and by constant (frequency independent) equilibrium distributions $n_{jk}^\text{eq}=n_j$ ($n_j\in\mathbb{C}$), see Eq.~(\ref{eq:Markov}). Following the ideas introduced in \cite{Lambert} for the Bosonic case, while some of the parameters in this formal model are allowed to take complex values, the dynamics is still defined using the same Schr\"odinger equation \emph{as if all parameters were physical}, i.e., without introducing any extra complex-conjugation, hence taking the name pseudo-Schr\"odinger equation \footnote{It has been written that the shortest and best way between two truths of the real domain often passes through the imaginary one \cite{Hadamard}.}. This allows us to use an \emph{unphysical} model to simulate a wider range of free correlation functions without adding extra complexity.  
To conclude its characterization, we assume the pseudo-environment to be initially in its equilibrium state  $\rho_{\text{pf}+\text{re}}$, see  Eq.~(\ref{eq:rho_pf}). 
\subsection{A Pseudo-Environment}
\label{app:pseudo-environment}
The Pseudo-Environment described by the Hamiltonian 
\begin{equation}
\label{eq:H_Spf}
    H_{S+\text{pf}}=H_S+H_\text{pf}+H^I_{S+\text{pf}}\;,
\end{equation}
is supposed to be, initially, in the following equilibrium state
\begin{equation}
\rho_{\text{pf}+\text{re}}=Z_{\text{pf}+\text{re}}^{-1}{\exp\displaystyle\sum_{j=1}^{N_\text{pf}}\left[\beta_j \Omega_j\tilde{c}_j^\dagger \tilde{c}_j+\sum_{k}\beta_{jk} \Omega_{jk}\tilde{c}^\dagger_{jk}\tilde{c}_{jk}\right]},
\end{equation}
in which $Z_{\text{pf}+\text{re}}$ ensures that the trace is $1$ and where each $\beta_j, \beta_{jk}\in\mathbb{R}$ can be found by imposing
\begin{equation*}
    \text{Tr}_\text{pf}[\tilde{c}^\dagger_{j}\tilde{c}_{j}\rho_\text{pf}]=n_j\;,
\end{equation*} 
on top of the already stated white-noise assumption
\begin{equation*}
    n_j^\text{eq}=\text{Tr}_\text{pf}[\tilde{c}^\dagger_{jk}\tilde{c}_{jk}\rho_\text{pf}]=n_j\;.
\end{equation*}
This leads to
\begin{equation}
\label{eq:constr_time_transl}
\beta_j\Omega_j=\beta_{jk}\Omega_{jk}=\log(1/n_j-1)\;,
\end{equation}
which concludes the characterization of this ``pseudo-environment''. We note that the condition in Eq.~(\ref{eq:constr_time_transl}) allows to write the exponent defining $\rho_{\text{pf}+\text{re}}$ as a weighted sum over the total number of Fermions in the different pseudo-environments, which is a constant of motion under the free evolution induced by $H_{\text{pf}+\text{re}}= H_\text{pf}+H^I_{\text{S}+\text{pf}}+H_{\text{re}}$ (as it only contains interaction terms which are written in a ``rotating wave'' style, i.e., they preserve the total number of bare excitations).
\subsection{Computing the correlation functions through the Heisemberg equation of motion}
\label{sec:pf_Heisemberg}
The most direct way to  compute the correlations in Eq.~(\ref{eq:corr_pf_free}) is through the Heisemberg equation of motion. For clarity, in this section we will be omitting the label $j$ used throughout the text to describe independent pseudo-environments.  In fact, we will be focusing on a single pseudo-fermion $\tilde{c}$ and its residual environment made of fermions $\tilde{c}_k$. With this notation, the free Hamiltonian of the environment, i.e., the part of the Hamiltonian in Eq.~(\ref{eq:H_Spf}) which has no support on the system, reads
 \begin{equation}
 \begin{array}{lll}
 H_{\text{pf}+\text{re}}&=&
 H_\text{pf}+H^I_{\text{S}+\text{pf}}+H_{\text{RE}}\\
 &=&\displaystyle\Omega \tilde{c}^\dagger \tilde{c}+\sum_k \lambda_k (\tilde{c}^\dagger \tilde{c}_k + \tilde{c}_k^\dagger \tilde{c}) +\sum_k\Omega_k \tilde{c}^\dagger_k \tilde{c}_k,
 \end{array}
 \end{equation}
 with $J(\omega)=\pi\sum_k \lambda_k^2\delta(\omega-\Omega_k)=\Gamma$. Using $\{\tilde{c},\tilde{c}^\dagger\}=1$ we obtain
 \begin{equation}
 \begin{array}{lll}
 \dot{\tilde{c}}&=&i[H^\text{pf}_E,d]=-i\Omega \tilde{c}-i\sum_k \lambda_k \tilde{c}_k\\
 \dot{c}_k&=&i[H^\text{pf}_E,\tilde{c}_k]=-i\Omega_k \tilde{c}_k-i\lambda_k\tilde{c}\;,
 \end{array}
 \end{equation}
 which, in Laplace space, becomes
  \begin{equation}
 \begin{array}{lll}
s\tilde{c}&=&\tilde{c}(0)-i\Omega \tilde{c}-i\sum_k \lambda_k \tilde{c}_k\\
s\tilde{c}_k&=&\tilde{c}_k(0)-i\Omega_k \tilde{c}_k-i\lambda_k \tilde{c}\;.
 \end{array}
 \end{equation}
Substituting 
\begin{equation*}
    \tilde{c}_k=(\tilde{c}_k(0)-i\lambda_k \tilde{c})/(s+i\Omega_k)
\end{equation*} 
into the first equation we get
\begin{equation}
\label{eq:temp_app_heis}
\left[s+i\Omega+\sum_k \frac{\lambda_k^2}{s+i\Omega_k}\right]\tilde{c}=\tilde{c}(0)-\sum_k\frac{i\lambda_k}{s+i\Omega_k}\tilde{c}_k(0).
\end{equation}
This equation can be made more explicit by computing
 \begin{equation}
 \label{eq:ii}
 \begin{array}{lll}
 \displaystyle\sum_k \frac{\lambda_k^2}{s+i\Omega_k}&=&\displaystyle\frac{1}{\pi}\int_{-\infty}^\infty d\omega \frac{J(\omega)}{s+i\omega}\\
 &=&\displaystyle\frac{\Gamma}{\pi}\int_{-\infty}^\infty  \frac{d\omega}{s+i\omega}\\
 &=&\displaystyle-i\frac{\Gamma}{\pi}\int_{-\infty}^\infty  \frac{d\omega}{\omega-is}.
 \end{array}
 \end{equation}
To proceed, we consider
 \begin{equation}
 \begin{array}{l}
\displaystyle \int_{-W}^W d\omega \frac{1}{\omega-is}
  =\displaystyle\log\left(\frac{|W-is|}{|-W-is|}\right)\\
  +\displaystyle\text{arg}(W-is)-\text{arg}(-W-is),
 \end{array}
 \end{equation}
 so that
  \begin{equation}
  \label{eq:W_W}
 \begin{array}{lllll}
\displaystyle\lim_{W\rightarrow\infty} \int_{-W}^W  \frac{d\omega}{\omega-is}&=&\displaystyle i\pi&\text{for}&\text{Re}(s)>0\\
&=&\displaystyle-i\pi&\text{for}& \text{Re}(s)<0\;.
 \end{array}
 \end{equation}
Which of these two alternative should we use in Eq.~(\ref{eq:temp_app_heis})? The choice depends on whether we are interested in evaluating the dynamics of $\tilde{c}(t)$ for $t>0$ or $t<0$. For $t>0$, we should chose the condition $\text{Re}(s)>0$ as it is always compatible with the integration path needed to define the inverse Laplace transform. For $t<0$, the opposite choice must be made. This means that
 \begin{equation}
 \label{eq:dir}
(s+i\Omega\pm\Gamma)\tilde{c}=\tilde{c}(0)-i\sum_k\frac{\lambda_k}{s+i\Omega_k}\tilde{c}_k(0)\;,
\end{equation}
where the $\pm$ depends on whether $t>0$ or $t<0$. A similar result holds for the conjugate
 \begin{equation}
 \label{eq:dir_conj}
(s-i\Omega\pm\Gamma)\tilde{c}=\tilde{c}^\dagger(0)+i\sum_k\frac{\lambda_k}{s-i\Omega_k}\tilde{c}^\dagger_k(0)\;.
\end{equation}
Note that, even if some of the original parameters $\Omega$ and $\Gamma$ might be complex, no complex conjugation is introduced on them, as a pseudo-Schr\"odinger equation is used to generate the dynamics.\\

Our goal is now to use Eq.~(\ref{eq:dir}) and Eq.~(\ref{eq:dir_conj}) to compute the correlations in Eq.~(\ref{eq:corr_pf_free}) which read
\begin{equation}
C^\sigma(t_2,t_1)=\lambda^2\text{Tr}[\tilde{c}^\sigma(t_2) \tilde{c}^{\bar{\sigma}}(t_1)\rho_{\text{pf}+\text{re}}]\;,
\end{equation}
where $\tilde{c}(t)=\exp\{iH^\text{pf}_Et\}\tilde{c}\exp\{-iH^\text{pf}_Et\}$ and where
\begin{equation}
\label{eq:rho_pf}
\rho_{\text{pf}+\text{re}}=\exp\{-\beta \Omega \tilde{c}^\dagger \tilde{c}\}\prod_k \exp\{-\beta_k\Omega_k \tilde{c}^\dagger_k \tilde{c}_k\}/Z_{\text{pf}+\text{re}}\;,
\end{equation}
where 
\begin{equation*}
    Z_{\text{pf}+\text{re}}=(e^{-\beta\Omega}+1)\prod_{k} (e^{-\beta_k\Omega_k}+1)\;,
\end{equation*}
together with the constraints in Eq.~(\ref{eq:constr_time_transl}), i.e., 
\begin{equation}
\label{eq:betadomegad}
\beta\Omega=\beta_k \Omega_k=\log(1/n-1)\;.
\end{equation}
Note that these constraints are essential to ensure invariance under time translation of the correlation. In fact, using them, the equilibrium state takes the form
\begin{equation*}
   \rho_{\text{pf}+\text{re}}\propto\exp\{-\beta\Omega N_{E}\}\;,
\end{equation*}
as a function of the total number of Fermions in the structured environment 
\begin{equation*}
  N_{\text{pf}+\text{re}}=\tilde{c}^\dagger \tilde{c}+\sum_k\tilde{c}^\dagger_k \tilde{c}_k\;.
\end{equation*}
This number is conserved by the free dynamics, i.e., $[H_\text{pf},N_\text{pf}]=0$ which implies that
\begin{equation}
\begin{array}{lll}
C^\sigma(t_2,t_1)&=&C^\sigma(t_2-t_1,0)\\
&\equiv&C^\sigma(t_2-t_1)\\
&=&\lambda^2\text{Tr}[\tilde{c}^\sigma(t) \tilde{c}^{\bar{\sigma}}(0)\rho_{\text{pf}+\text{re}}]\;,
\end{array}
\end{equation}
where $t=t_2-t_1$. Therefore, using Eq.~(\ref{eq:dir}) and Eq.~(\ref{eq:dir_conj}), we can immediately compute
\begin{equation}
\begin{array}{lll}
C^\sigma(t)&=&\displaystyle\lambda^2\theta(t)\mathcal{L}_t^{-1}\frac{\text{Tr}[\tilde{c}^\sigma(0) \tilde{c}^{\bar{\sigma}}(0) \rho_{\text{pf}+\text{re}}]}{s-\sigma i\Omega+\Gamma}\\
&&\displaystyle+\lambda^2\theta(-t)\mathcal{L}_t^{-1}\frac{\text{Tr}[\tilde{c}^\sigma(0) \tilde{c}^{\bar{\sigma}}(0) \rho_{\text{pf}+\text{re}}]}{s-\sigma i\Omega-\Gamma}\;,
\end{array}
\end{equation}
where $\mathcal{L}_t^{-1}$ is the inverse Laplace transform which can be computed to obtain
\begin{equation}
C(t)= \lambda^2 [{(1-\sigma)}/{2}+\sigma n]\exp\{\sigma i\Omega t -\Gamma |t|\},
\end{equation}
which, reintroducing the indexes $j$ used in the main text to label independent pseudo-environments, leads to Eq.~(\ref{eq:corr_pf_pf}) by linearity.

\subsection{Markovian Regime}
In this section, we review the limit in which the environment is Markovian. This regime is defined when all the memory effects described by the Fermionic influence superoperator $\mathcal{F}(t,s,C^\sigma)$ in Eq.~(\ref{eq:rhoF}) are negligible. The Markovian limit can be recovered \cite{Cirio2021,Gardiner} in the quantum white-noise limit characterized by a constant spectral density and a constant equilibrium distribution, i.e.,
\begin{equation}
\label{eq:Markov}
    J(\omega)\mapsto\Gamma_0,\qquad
    n_E^\text{eq}(\omega)\mapsto n_0\;.
\end{equation}
In fact, these conditions, once inserted in Eq.~(\ref{eq:corr_short}) lead to 
\begin{equation*}
    C^\sigma(t_2,t_1)=\Gamma_0 [1-\sigma+2\sigma n_0]\delta(t_2-t_1)\;,
\end{equation*}
which, arguably, more commonly defines the Markov approximation. As shown in \cite{Cirio2021}, in the case of a delta-correlated environment, the expression for the functional superoperator in Eq.~(\ref{eq:rhoF}) drastically simplifies, leading to an effective generalized Lindblad equation of motion
\begin{equation}
\label{eq:Lindblad}
    \begin{array}{lll}
         \displaystyle\rho_S&=&\displaystyle-i[H_S,\rho_S]+\Gamma_0\sum_{r=\pm} (1-n_0)D_s^r[\rho_S^r]+n_0 D_{s^\dagger}^r[\rho_S^r]
    \end{array}
\end{equation}
 where $\rho^{r=\pm}_S$, are the projections of the density matrix into the space with even/odd Fermionic parity. The dissipators $D_O^r[\cdot]$ are defined as
 \begin{equation*}
     D_O^r[\cdot]=2rO[\cdot]O^\dagger-O^\dagger O[\cdot]-[\cdot]O^\dagger O\;.
 \end{equation*}
 
In non-Markovian regimes, the influence superoperator $\mathcal{F}$ is able to model more complex physical scenarios. For example, in the limit where the system is a single-level impurity one recovers the single-impurity Anderson model, which is integrable in the non-interacting limit.  Beyond this case, with e.g., interacting or non-linear impurities, one must resort to numerical methods.

This analysis can be easily adapted to the problem of tracing out the residual environments of the pseudofermion model defined in Eq.~(\ref{eq:H_Spf}). In fact, 
the residual environment of each pseudo-fermion is modeled as idealized white noise and, as a consequence, it can be traced out by simply 
considering the composite system+pseudofermions in place of $S$ in Eq.~(\ref{eq:Lindblad}) leading to Eq.~(\ref{eq:master_eq_pf}).
\subsection{Correlations for a Lorentzian spectral density}
\label{app:corr_lorentzian}
We now consider the Lorentzian spectral density
\begin{equation}
J_L(\omega)=\frac{\Gamma W^2}{[(\omega-\omega_0)^2+W^2]}=\frac{\Gamma W^2}{(\omega-a)(\omega-\bar{a})},
\end{equation}
where $a=\omega_0+iW$ and $\bar{a}=\omega_0-iW$. With this spectral density, the correlation functions in Eq.~(\ref{eq:corr_short}) take the form
\begin{equation}
\begin{array}{lll}
C^{\sigma=1}(t)&=&\displaystyle\int_{-\infty}^\infty \frac{d\omega}{\pi} J_L(\omega) e^{i\omega t} n_{\beta\mu}(\omega)\\
C^{\sigma=-1}(t)&=&\displaystyle\int_{-\infty}^\infty \frac{d\omega}{\pi} J_L(\omega) e^{-i\omega t} [1-n_{\beta\mu}(\omega)],
\end{array}
\end{equation}
and can be evaluated by noticing that the poles of the integrand are located at $a$, $\bar{a}$, and $\omega_k=\mu+i x_k$,  where $x_k=(2k-1)i\pi/\beta$, for $k\in\mathbb{N}$.
For positive (negative) time arguments, we can close the contour in the upper (lower) complex plane to derive, for $t>0$
\begin{equation}
\label{eq:corr_ana}
\begin{array}{l}
C^{\sigma=1}(t)=\displaystyle 2i\Gamma W^2\left[\frac{e^{i at}}{a-\bar{a}}\frac{1}{e^{\beta(a-\mu)}+1}\right.\\
\phantom{C^{\sigma=1}(t)}-\displaystyle\left.\frac{1}{\beta}\sum_{k>0}\frac{e^{i\omega_k t}}{(\omega_k-a)(\omega_k-\bar{a})}\right]\\
C^{\sigma=1}(-t)=\displaystyle 2i\Gamma W^2\left[\frac{e^{-i \bar{a}t}}{a-\bar{a}}\frac{1}{e^{\beta(\bar{a}-\mu)}+1}\right.\\
\phantom{C^{\sigma=1}(-t)}+\displaystyle\left.\frac{1}{\beta}\sum_{k> 0}\frac{e^{-2i\mu t}e^{i\omega_k t}}{(2\mu-\omega_k-a)(2\mu-\omega_k-\bar{a})}\right]\\

C^{\sigma=-1}(t)=\displaystyle -2i\Gamma W^2\left[\frac{e^{-i \bar{a} t}}{\bar{a}-a}\frac{e^{\beta(\bar{a}-\mu)}}{e^{\beta(\bar{a}-\mu)}+1}\right.\\
\phantom{C^{\sigma=-1}(t)}+\displaystyle\left.\frac{1}{\beta}\sum_{k>0}\frac{e^{-2i\mu t}e^{i\omega_k t}}{(2\mu-\omega_k-a)(2\mu-\omega_k-\bar{a})}\right]\\

C^{\sigma=-1}(-t)=\displaystyle  2i\Gamma W^2\left[\frac{e^{-\beta (\mu-a)}e^{i {a}t}}{a-\bar{a}}\frac{1}{e^{\beta({a}-\mu)}+1}\right.\\
\phantom{C^{\sigma=-1}(-t)}+\displaystyle\left.\frac{1}{\beta}\sum_{k>0}\frac{e^{i\omega_k t}}{(\omega_k-a)(\omega_k-\bar{a})}\right],
\end{array}
\end{equation}
Noticing that $\bar{\omega}_k=2\mu-\omega_k$, the above expressions are compatible with the general relation $C^\sigma(-t)=\bar{C}^\sigma(t)$.
\begin{equation}
\label{eq:qwe}
\begin{array}{l}
C^{\sigma=-1}(-t)=
\displaystyle  2i\Gamma W^2\left[\frac{e^{i {a}t}}{a-\bar{a}}\left(1-\frac{1}{e^{\beta({a}-\mu)}+1}\right)\right.\\
\phantom{C^{\sigma=-1}(-t)=}+\displaystyle\left.\frac{1}{\beta}\sum_{k>0}\frac{e^{i\omega_k t}}{(\omega_k-a)(\omega_k-\bar{a})}\right]\\

\phantom{C^{\sigma=-1}(-t)}=\displaystyle  2i\Gamma W^2\frac{e^{i {a}t}}{a-\bar{a}}-C^{\sigma=1}(t).
\end{array}
\end{equation}
Using the Matsubara expansion (see \cite{Mahan}, Eq.~(3.5), pag.~110)
\begin{equation}
\frac{1}{e^{\beta x}+1}=\frac{1}{2}-\frac{1}{\beta}\sum_{k>0}\left(\frac{1}{x-ix_k}+\frac{1}{x+ix_k}\right)
\end{equation}
we can alternatively write, for $t>0$,
\begin{equation}
\begin{array}{l}
C^{\sigma=1}(t)=\\
=\displaystyle 2i\Gamma W^2\left\{\frac{e^{i a t}}{a-\bar{a}}\left[\frac{1}{2}-\frac{1}{\beta}\sum_{k>0}\left(\frac{1}{a-\omega_k}\right.\right.\right.\\
\displaystyle\phantom{=}\left.\left.+\frac{1}{a-2\mu+\omega_k}\right)\right]\left.-\frac{1}{\beta}\sum_{k>0}\frac{e^{i\omega_k t}}{(\omega_k-a)(\omega_k-\bar{a})}\right\}\\

=\displaystyle \left\{\frac{e^{i a t}}{a-\bar{a}}\left[\frac{1}{2}-\frac{1}{\beta}\sum_{k>0}\left(\frac{1}{a-\omega_k}+\frac{1}{\omega_k-\bar{a}}\right)\right]\right.\\
\displaystyle\phantom{=}\left.-\frac{1}{\beta}\sum_{k>0}\frac{e^{i\omega_k t}}{(\omega_k-a)(\omega_k-\bar{a})}\right\}2i\Gamma W^2\\

=\displaystyle 2i\Gamma W^2\left\{\frac{e^{i a t}}{2(a-\bar{a})}-\frac{1}{\beta}\sum_{k>0}\frac{e^{i\omega_k t}-e^{ia t}}{(\omega_k-a)(\omega_k-\bar{a})}\right\}\\

=\displaystyle \frac{\Gamma W}{2}e^{i \mu t-Wt}+\frac{2i\Gamma W^2}{\beta}\sum_{k>0}\frac{e^{i\mu t-x_k t}-e^{i\mu t-W t}}{(x_k-W)(x_k+W)},
\end{array}
\end{equation}
where we noticed that $1/(a-2\mu+\omega_k)-1/(\omega_k-\bar{a})=0$, $\forall k$.
Using $C^\sigma(-t)=\bar{C}^\sigma(t)$, together with Eq.~(\ref{eq:qwe}), we can write, for $t>0$,
\begin{equation}
\begin{array}{l}
C^{\sigma=-1}(t)=\bar{C}^{\sigma=-1}(-t)\\
=\displaystyle  \Gamma W e^{-i\mu t-W t}-\bar{C}^{\sigma=1}(t)\\

=\displaystyle \frac{\Gamma W}{2}e^{-i \mu t-Wt}+\frac{2i\Gamma W^2}{\beta}\sum_{k>0}\frac{e^{-i\mu t-x_k t}-e^{-i\mu t-W t}}{(x_k-W)(x_k+W)}.
\end{array}
\end{equation}
Using $C^\sigma(-t)=\bar{C}^\sigma(t)$ again, we can extend these results to $t<0$ to write, for any $t\in\mathbb{R}$
\begin{equation}
C_L^\sigma(t)=C^\sigma_\text{res}+\sum_{k>0}M_k^\sigma(t)\;,
\end{equation}
where
\begin{equation}
\label{eq:akn}
\begin{array}{lll}
C^\sigma_\text{res}&=&\displaystyle\frac{\Gamma W}{2}\exp\{\sigma i \mu t-W|t|\}\\
M_k^\sigma(t)&=&\displaystyle \frac{\text{sg}(t)2i\Gamma W^2}{\beta}\frac{e^{\sigma i \mu t-x_k |t|}-e^{\sigma i \mu t-W |t|}}{x^2_k-W^2}\;,
\end{array}
\end{equation}
where $\text{sg}(t)=t/|t|$ for $t\neq 0$, is the sign function.
One interesting feature of this decomposition is that the pole at $x_k=W$ is explicitly removed from the notation (because of the presence of a corresponding zero in the numerator). The first line of Eq.~(\ref{eq:akn}) reproduces the first line of Eq.~(\ref{eq:res_and_Mats}) in the main text. In order to reproduce the second line of Eq.~(\ref{eq:res_and_Mats}) there is a little more work to do.
To achieve this, we use the identity in Eq.~(\ref{eq:sign_identity}) to write
\begin{equation}
\begin{array}{l}
\text{sg}(t)(e^{-x_k|t|}-e^{-W|t|})=\\
\displaystyle=\frac{e^{\omega(t+|t|)}-e^{\omega(|t|-t)}}{e^{2\omega|t|}-1}e^{-W|t|}(e^{-(x_k-W)|t|}-1),
\end{array}
\end{equation}
for any $\omega\in\mathbb{C}$. For the specific choice $\omega=(W-x_k)/2$ we obtain
\begin{equation}
\begin{array}{l}
\text{sg}(t)(e^{-x|t|}-e^{-W|t|})=\\
=\displaystyle e^{-(W+x)|t|/2} [e^{(W-x)t/2}-e^{-(W-x)t/2}],
\end{array}
\end{equation}
which, used in the second line of Eq.~(\ref{eq:akn}) gives
\begin{equation}
\label{eq:M_k}
\begin{array}{l}
M^{\sigma}_k(t)=\displaystyle \sum_{k>0}\frac{2i\Gamma W^2}{\beta}\frac{e^{-(W+x_k)|t|/2}}{{(x^2_k-W^2)}}\\
\displaystyle\times\left[{e^{[i\sigma\mu+(W-x_k)/2]t}-e^{[i\sigma\mu-(W-x_k)/2]t}}\right]\;,
\end{array}
\end{equation}
which is the second line in Eq.~(\ref{eq:res_and_Mats}).

We finish noting that, in the zero-temperature limit ($\beta\rightarrow\infty$), the Matsubara frequencies $x_k$ approach a continuum so that
\begin{equation}
\label{eq:Matsubara_integral}
\begin{array}{lll}
M^{\sigma}(t)&\!\!\!\!\overset{\beta\rightarrow\infty}{=}\!\!\!\!\!\!&\displaystyle \text{sg}(t)\frac{i\Gamma W^2}{\pi}{e^{i\sigma\mu t}}\int_0^\infty \!\!\!\!dx\frac{e^{-x |t|}-e^{-W|t|}}{(x^2-W^2)}.
\end{array}
\end{equation}
\subsection{Correspondence to pseudo-environments}
Here we provide details about modeling the correlations $C^{\sigma}_\text{res}(t)$ and
$M_k^{\sigma}(t)$ in Eq.~(\ref{eq:res_and_Mats}) using Fermionic pseudo-environments. We start from the resonant contribution $C^{\sigma}_\text{res}(t)$. We want to find the parameters of a pseudo-environment such that its free correlation function $C^\sigma_{\text{pf},\text{res}}(t)$, obtained using the identification $j\mapsto\text{res}$ in Eq.~(\ref{eq:corr_pf_pf}), fulfills 
\begin{equation}
\begin{array}{lll}
C^{\sigma}_\text{res}(t)&=&C^\sigma_{\text{pf},\text{res}}(t)\;.
\end{array}
\end{equation}
Using Eq.~(\ref{eq:res_and_Mats})  and Eq.~(\ref{eq:corr_pf_pf}) the equation above translates to  
\begin{equation}
\label{eq:1:1}
\begin{array}{lll}
\displaystyle \frac{\Gamma W}{2} e^{i \sigma\mu t - W |t|}&=&\lambda_\text{res}^2[(1-\sigma)/2+\sigma n_\text{res}]e^{i\sigma\Omega_\text{res}t-\Gamma_\text{res} |t|}.
\end{array}
\end{equation}
The equivalence in Eq.~(\ref{eq:1:1}) can be imposed by defining
\begin{equation}
\begin{array}{lll}
n_\text{res}&=&1/2\\
\lambda_\text{res}&=&\sqrt{\Gamma W/(2 n_\text{res})} \\
\Omega_\text{res}&=&\mu\\
\Gamma_\text{res}&=&W\;,
\end{array}
\end{equation}
which fully characterize the resonant pseudo-environment.
Similarly, for each $k$, the Matsubara contribution $M_k^\sigma(t)$ in  Eq.~(\ref{eq:res_and_Mats}) can be reproduced using two pseudo-environments. Explicitly, identifying $j\rightarrow (k,r)$, $r=\pm$ in  Eq.~(\ref{eq:corr_pf_pf}), we want to impose
\begin{equation}
\label{eq:temppp}
\begin{array}{lll}
M^\sigma_{k}(t)&=&\displaystyle\sum_{r=\pm}C^{\sigma}_{\text{pf},(k,r)}(t)\;,
\end{array}
\end{equation}
which, using Eq.~(\ref{eq:res_and_Mats})  and Eq.~(\ref{eq:corr_pf_pf}) is equivalent to 
\begin{equation}
\label{eq:1:2}
\begin{array}{lll}
\displaystyle M_k e^{-(W+x_k)|t|/2}\sum_{r=\pm}r\exp\{[i\sigma\mu+r(W-x_k)/2]t\}\\
=\displaystyle\sum_{r=\pm}\lambda_{k,r}^2[(1-\sigma)/2+\sigma n_{k,r}]\exp\{i\sigma\Omega_{k,r}t-\Gamma_{k,r} |t|\}
\end{array}
\end{equation}
Imposing the equation above requires a bit of attention as the frequency $r(W-x_k)/2$ appearing in the expression of $M_k^\sigma(t)$ is \emph{not} multiplied by the parameter $\sigma$ as in the correlation for the pseudo-environment. On the contrary, the coefficients multiplying the exponential in the correlation for the pseudo-environment do depend on $\sigma$ while $M_k$ does not. Luckily, the Matsubara contributions $M_k^\sigma(t)$ are written in terms of a difference between exponentials with opposite frequencies which offers the opportunity to define the parameters characterizing the Matsubara pseudo-environments as
\begin{equation}
\begin{array}{lll}
n_{k,\pm}&=&\Delta\\
\lambda_{k,\pm}&=&\sqrt{ \pm M_k / \Delta } \\
\Omega_{k,\pm}&=&\mu\mp i(x_k-W)/2\\
\Gamma_{k,\pm}&=&(W+x_k)/2\;.
\end{array}
\end{equation}
where $\Delta\in\mathbb{C}$ in the limit $|\Delta|\rightarrow\infty$ so that
\begin{equation}
\begin{array}{lll}
C^{\sigma}_{\text{pf},(k,r)}(t)
&=&\displaystyle\frac{ r M_k}{\Delta}e^{-(W+x_k)|t|/2}[(1-\sigma)/2+\sigma \Delta]\\
&&\displaystyle\times \exp\{i\sigma[\mu- ir(x_k-W)/2]t\}\;,
\end{array}
\end{equation}
which, inserted in Eq.~(\ref{eq:temppp}), leads to
\begin{equation}
\begin{array}{lll}
M_k^\sigma(t)
&=&\displaystyle\frac{ M_k}{\Delta}e^{-(W+x_k)|t|/2}[(1-\sigma)/2+\sigma \Delta]\\
&&\displaystyle\times \left(e^{i\sigma[\mu- i(x_k-W)/2]t}-e^{i\sigma[\mu+i(x_k-W)/2]t}\right)\\

&=&\displaystyle\frac{ M_k}{\Delta}e^{-(W+x_k)|t|/2}[(1-\sigma)/2+\sigma \Delta]\\
&&\displaystyle\times  e^{i\sigma\mu t}\left(e^{\sigma[(x_k-W)/2]t}-e^{-\sigma[(x_k-W)/2]t}\right)\\

&=&\displaystyle\frac{ M_k}{\Delta}e^{-(W+x_k)|t|/2}[(1-\sigma)/2+\sigma \Delta]\\
&&\displaystyle\times  \sigma e^{i\sigma\mu t}\left(e^{[(x_k-W)/2]t}-e^{-[(x_k-W)/2]t}\right)\\

&\rightarrow&\displaystyle M_ke^{-(W+x_k)|t|/2}\\
&&\displaystyle\times  e^{i\sigma\mu t}\left(e^{[(x_k-W)/2]t}-e^{-[(x_k-W)/2]t}\right)\\
\end{array}
\end{equation}
where, in the last step, the limit $|\Delta|\rightarrow\infty$ was taken. The above equation is equivalent to the expression for $M_k^\sigma(t)$ given in Eq.~(\ref{eq:res_and_Mats}), thereby completing the proof. In numerical applications, the limit $\Delta\rightarrow\infty$ might introduce some numerical instabilities which can be regularized using intermediate values such that $\Delta\gg 1$.\\

Interestingly, it is also possible to build a pseudo-environment which does not resort to any asymptotic parameter ($\Delta$ above). To achieve this, we need to introduce four pseudo-environments to model each $M_k^\sigma(t)$ to impose
\begin{equation}
\label{eq:tempppp}
\begin{array}{lll}
M^\sigma_{k}(t)&=&\displaystyle\sum_{r,\sigma'=\pm}C^{\sigma}_{\text{pf},(k,r,\sigma')}(t)\;,
\end{array}
\end{equation}
through the identification  $j\rightarrow (k,r,\sigma')$, $r,\sigma'=\pm$ in  Eq.~(\ref{eq:corr_pf_pf}). In order to fulfill Eq.~(\ref{eq:tempppp}), we can choose the parameters
\begin{equation}
\begin{array}{lll}
n_{k,r,\sigma'}&=&(1+\sigma')/2\\
\lambda_{k,r,\sigma'}&=&\sqrt{r M_k} \\
\Omega_{k,r,\sigma'}&=&\mu - i r\sigma'(x_k-W)/2\\
\Gamma_{k,r,\sigma'}&=&(W+x_k)/2\;,
\end{array}
\end{equation}
which correspond to a $\beta\rightarrow\sigma'\times\infty$ limit, so that, see Eq.~(\ref{eq:corr_pf_pf}),
\begin{equation}
\begin{array}{lll}
C^{\sigma}_{\text{pf},(k,r,\sigma')}(t)
&=&\displaystyle r M_ke^{-(W+x_k)|t|/2}\left[\frac{1-\sigma}{2}+\sigma \frac{1+\sigma'}{2}\right]\\
&&\displaystyle\times e^{i\sigma[\mu- ir\sigma'(x_k-W)/2]t}\\
\end{array}
\end{equation}
which, inserted in Eq.~(\ref{eq:tempppp}), leads to
\begin{equation}
\begin{array}{lll}
M_k^\sigma(t)
&=&\displaystyle\sum_{\sigma'} M_ke^{-(W+x_k)|t|/2}\left[\frac{1-\sigma}{2}+\sigma \frac{1+\sigma'}{2}\right]\\
&&\displaystyle\times \left(e^{i\sigma[\mu- i\sigma'(x_k-W)/2]t}-e^{i\sigma[\mu+i\sigma'(x_k-W)/2]t}\right)\\

&=&\displaystyle\sum_{\sigma'} M_ke^{-(W+x_k)|t|/2}\delta_{\sigma\sigma'}\\
&&\displaystyle\times  e^{i\sigma\mu t}\left(e^{\sigma\sigma'(x_k-W)t/2}-e^{-\sigma\sigma'(x_k-W)t/2}\right)\\

&=&\displaystyle M_ke^{-(W+x_k)|t|/2}\\
&&\displaystyle\times  e^{i\sigma\mu t}\left(e^{(x_k-W)t/2}-e^{-(x_k-W)t/2}\right),
\end{array}
\end{equation}
which, similarly as in the previous analysis, is equivalent to the expression for $M_k^\sigma(t)$ given in Eq.~(\ref{eq:res_and_Mats}), thereby completing the proof. \\
\subsection{Proof of an identity for the sign function}
\label{app:proof_Mk}
Let us define $\text{sg}(x)=x/|x|$ for $x\neq 0$. We have
\begin{equation}
\begin{array}{lll}
e^{\omega(t+|t|)}-e^{\omega(|t|-t)}&\overset{t>0}{=}&e^{2\omega t}-1\\
&=&e^{2\omega|t|}-1\\
&\overset{t<0}{=}&1-e^{-2\omega t}\\
&=&1-e^{2\omega|t|}\;,
\end{array}
\end{equation}
for any $\omega\in \mathbb{C}$. In a more compact notation, the previous equation can be written as
\begin{equation}
\label{eq:sign_identity}
\begin{array}{lll}
e^{\omega(t+|t|)}-e^{\omega(|t|-t)}=\text{sg}(t)(e^{2\omega|t|}-1)\;.
\end{array}
\end{equation}

\section{Effective model for fast decaying terms in the Matsubara correlation}
For Bosonic environments, when a term in the Matsubara series has a decay rate which is larger than the highest frequency $\Omega_S$ which can be associated to the system, it is possible to approximate its effect on the system by adding an extra dissipator to the master equation \cite{Ishizaki_1}. In this section, we analyze the same limit in the case of a Fermionic environment interacting with the system through the Lorentzian spectral density $J_L(\omega)$. In this case, the Matsubara series takes the form described in Eq.~(\ref{eq:res_and_Mats}), i.e., 
\begin{equation}
\begin{array}{lll}
M^{\sigma}(t)&=&\displaystyle \text{sg}(t)\frac{2i\Gamma W^2}{\beta}{e^{i\sigma\mu t}}\sum_{k>0}\frac{e^{-x_k |t|}-e^{-W|t|}}{(x^2_k-W^2)},
\end{array}
\end{equation}
where $x_k=(2k-1)\pi/\beta$. As in the Bosonic case, our starting point is the following limit-representation of the Dirac delta \cite{Ishizaki_1}
\begin{equation}
\delta(t)=\lim_{\epsilon\rightarrow 0}\frac{e^{-|t|/\epsilon}}{2\epsilon}\;.
\end{equation}
Among the exponentials present in the Matsubara series above, the ones which can be modeled with a delta contribution are those for which either $x_k\gg\Omega_S$ or $W\gg \Omega_S$. We note that the former possibility corresponds to a ``high-temperature'' limit (as it corresponds to $\beta\Omega_S\ll (2k-1)\pi$); while the latter corresponds to a ``broad spectral density'' limit. Whenever we are in one of these regimes, the corresponding contribution in the Matsubara series can be approximated as
\begin{equation}
\begin{array}{lll}
M^\sigma_\text{eff}(t)&=&\text{sg(t)}A_\text{eff}\delta(t)\;,
\end{array}
\end{equation}
where $A_\text{eff}=i\Gamma_\text{eff}$ in terms of an effective decay rate $\Gamma_\text{eff}$. We can now compute the corresponding contribution to the influence superoperator, i.e., we can compute Eq.~(\ref{eq:rhoF}) with the replacement $C^\sigma(t_2,t_1)\mapsto M_\text{eff}(t_2-t_1)$. We find
\begin{equation}
\mathcal{F}(t)=\int_0^t d t_2\int_0^{t_2}dt_1\;\mathcal{W}(t_2,t_1)\;,
\end{equation}
where $\mathcal{W}(t_2,t_1)=\sum_{\sigma=\pm}\mathcal{A}^\sigma(t_2) \mathcal{B}_\text{eff}^\sigma(t_2,t_1)$ with
\begin{equation*}
\begin{array}{lll}
\mathcal{A}^\sigma(t)[\cdot]&=&\displaystyle\hat{s}^{\bar{\sigma}}(t)[\cdot]-\mathcal{P}_S[[\cdot]\hat{s}^{\bar{\sigma}}(t)]\\
\mathcal{B}_\text{eff}^{\sigma}(t_2,t_1)[\cdot]&=&-M_\text{eff}^{\sigma}(t)\hat{s}^\sigma(t_1)\cdot-\bar{M}^{\bar{\sigma}}_\text{eff}(t)\mathcal{P}_S[[\cdot]\hat{s}^\sigma(t_1)]\\
&=&-A_\text{eff}\delta(t)\hat{s}^\sigma(t_1)\cdot-\bar{A}_\text{eff}\delta(t)\mathcal{P}_S[[\cdot]\hat{s}^\sigma(t_1)],
\end{array}
\end{equation*}
where $t=t_2-t_1\geq 0$, which justifies the last step. We then obtain
\begin{equation*}
\label{eq:temp_L}
\begin{array}{lll}
W(t_2,t_1)&=&-\displaystyle\delta(t)(s\cdot-\mathcal{P}_S[\cdot s])(A_\text{eff} s^\dagger\cdot+\bar{A}_\text{eff}\mathcal{P}_S[\cdot s^\dagger])/2\\
&&-\displaystyle\delta(t)(s^\dagger\cdot-\mathcal{P}_S[\cdot s^\dagger])(A_\text{eff} s\cdot+\bar{A}_\text{eff}\mathcal{P}_S[\cdot s])/2.\\
\end{array}
\end{equation*}
This means that, in the even/odd sector, in the Schr\"odinger picture, we have
\begin{equation*}
\begin{array}{lll}
\mathcal{F}(t)&=&-t\displaystyle A_\text{eff} (ss^\dagger\cdot-\cdot s^\dagger s\pm s\cdot s^\dagger\mp s^\dagger\cdot s)/2\\
&&-t\displaystyle A_\text{eff}(s^\dagger s\cdot-\cdot s s^\dagger\pm s^\dagger\cdot s\mp s\cdot s^\dagger)/2\\
&=&-A_\text{eff}/2([ss^\dagger+s^\dagger s,\cdot]) t\;,
\end{array}
\end{equation*}
which, if $s$ is such to satisfy the Fermionic anticommutation rules, is zero. 

In conclusion, when a term in the Matsubara correlation function in Eq.~(\ref{eq:res_and_Mats}) can be modeled as a delta function, it does not bring any effect on the system dynamics. For example, this implies that the Matsubara correlation function can be neglected when both the following conditions are satisfied: the high temperature limit (i.e., $1/\beta$ much bigger than the highest frequency associated with the system) and a wide Lorentzian spectral density (i.e., with a width $W$ much bigger than the highest frequency associated with the system).

\section{Tensor Network Simulation of the Pseudofermion Lindblad Master Equation}

\begin{figure}[t!]
\includegraphics[width = \columnwidth]{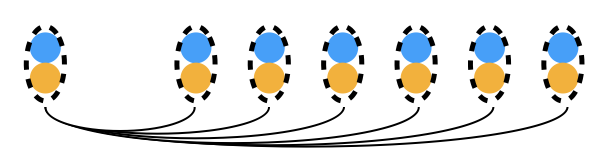}
\caption{Schematics for the superfermion representation. Blue circles correspond to physical fermions and orange circles auxiliary ones. They are combined in pairs to give one MPS site, shown as dashed ellipses. The first one comes from the system while others from the bath. Solid lines below indicate the long-ranged hopping terms in the Hamiltonian. }
\label{fig:MPSschematics}
\end{figure}

Here we describe how to simulate Eq.~(\ref{eq:master_eq_pf}) of the main text with tensor networks. In the algorithm we have used the superfermion representation of the master equation and the Swap-gate technique to treat long-ranged hopping which arises due to the use of energy eigenbasis for the environments. Since we closely follow Ref.~\cite{PhysRevX.10.031040}, we only elaborate some key aspects in designing the algorithm instead of presenting all details here. 

The superfermion representation is a way to map the fermionic master equation onto a non-Hermitian Schr{\"o}dinger equation, similiar to the purification technique of simulating master equations \cite{SCHOLLWOCK201196}. In the superfermion representation, an extra auxiliary fermion is introduced to each physical fermion, doubling the dimension of the Hilbert space. The arrangement of physical and auxiliary fermions is in principle arbitrary, but for the purpose of tensor network simulation, it is beneficial to arrange them in a intertwined order, see Fig.~\ref{fig:MPSschematics}. 

A key object in the construction is the so-called left-vacuum state defined by
\begin{equation}
|I\rangle = |0\underline0,0\underline0,\cdots,0\underline0\rangle\;,
\end{equation}
where $0$ $(1)$ stands for empty (filled) fermionic site and we have combined the physical fermion and the corresponding auxiliary one (marked by underlines) into pairs. By applying the density matrix and master equation to $|I\rangle$ from the left and making use of the following conjugation rules
\begin{equation}
d^\dagger |I\rangle = - \underline{d} |I\rangle,\; d|I\rangle = \underline{d}^\dagger |I\rangle 
\end{equation}
where $d$ indicates system fermions ($s$) and pseudo-fermions ($\tilde c$) and $\underline{d}$ its auxiliary counterpart, we arrive at a non-Hermitian Schr{\"o}dinger equation 
\begin{equation}
\label{eq:superfschroedinger}
d|\rho\rangle/dt = -i L|\rho\rangle    
\end{equation}
with $|\rho\rangle=\rho|I\rangle$ and 
\begin{equation}
    \begin{array}{lll}
L &=& H_{S_\text{pf}}^{(0)} - \underline{H}_{S+\text{pf}}^{(0)} \nonumber\\
&&+ \sum_j \Bigg[ \Gamma_jn_j\Big(\tilde{c}_j^\dagger \underline{\tilde{c}}_j^\dagger-\frac12(\tilde{c}_j\tilde{c}_j^\dagger+\underline{\tilde{c}}_j\underline{\tilde{c}}_j^\dagger)\Big) \nonumber\\
&&-\Gamma_j(1-n_j)\Big(\tilde{c}_j \underline{\tilde{c}}_j-\frac12(\tilde{c}_j^\dagger\tilde{c}_j+\underline{\tilde{c}}_j^\dagger\underline{\tilde{c}}_j)\Big) \Bigg].
\end{array}
\end{equation}
Note that Eq.~(\ref{eq:superfschroedinger}) has no explicit dependence on the fermion parity, which is one of the main advantages in taking the superfermion representation. 

For the tensor network simulation, it is more efficient to combine a physical fermion and its partner into one MPS site, so that the physical index has dimension four in our simulation. One immediately sees in this way that the Lindblad terms in $L$ are all local, which act on a single MPS site; Hence easy to be handled in the simulation. 

The hopping terms in the Hamiltonian are between the pseudofermions and the system, which constitutes the so-called star-geometry, needs some special care. In our implementation we use the Swap gate 
\begin{equation}\label{eq:MPSswap}
    \mathcal{S} = (I_2\otimes S \otimes I_2)(S\otimes S)(I_2\otimes S \otimes I_2)
\end{equation}
to exchange the local states of nearest-neighbour MPS sites, here $I_2$ is the $2\times2$ identity matrix and $S$ can be explicitly written as 
\begin{eqnarray}
S = 
\begin{pmatrix}
1 & 0 & 0 & 0 \\
0 & 0 & 1 & 0 \\
0 & 1 & 0 & 0 \\
0 & 0 & 0 & -1
\end{pmatrix}, 
\end{eqnarray}
in the basis made by $\{|0\underline0\rangle,|0\underline1\rangle,|1\underline0\rangle,|1\underline1\rangle\}$. Intuitively, $S$ interchanges the states of two adjacent fermionic sites, while $\mathcal{S}$ swaps the states of two adjacent MPS sites (each consisting of two fermionic sites). For example, by labeling an MPS state as $|n_1\underline {n}_1, n_2\underline{n}_2\rangle$, the operator $I_2\otimes S\otimes I_2$ swaps $\underline{n}_1\leftrightarrow n_2$, the operator $S\otimes S$ swaps $n_1\leftrightarrow n_2$ and $\underline{n}_1\leftrightarrow\underline{n}_2$, and the operator $I_2\otimes S\otimes I_2$ swaps $n_1\leftrightarrow\underline{n}_2$, resulting in the expression presented in Eq.~(\ref{eq:MPSswap}).

Therefore any long-ranged two-site gate can be decomposed into a sequence of Swap gates and nearest-neighbour gates, which can be implemented in tensor networks efficiently. We have used a second-order Trotter decomposition of the propagator and by carefully choosing the order of gates, many Swap gates arising due to long-range hopping can be merged into the identity operator, thus avoiding numerical overheads. 

\newpage

\nocite{apsrev41Control}
\bibliographystyle{apsrev4-2}

\bibliography{bib}

\end{document}